\documentclass[12pt]{article}
\usepackage{graphicx}
\usepackage{ amssymb }
\usepackage{bm}
 \oddsidemargin  = - 7 mm 
 \evensidemargin = - 7 mm
\def\jtstrut{\vrule height4ex depth0pt width0pt} 

\newcommand{\bD}{{\bar D}}
\newcommand{\MeV}{{\rm MeV}}

\newcommand{\ignore}[1]{}

\newcommand{\uno}{{\bf 1}}

\newcommand{\sesentaytres}{{\bf 63}}

\begin{document}

\title{$D^-$ mesic atoms} 
\author{
 {\large C. Garc\'{\i}a-Recio
}\\
 {\small Departamento de F\'{\i}sica At\'omica Molecular y Nuclear, 
  and} \\ {\small
Instituto Carlos I de F{\'\i}sica Te\'orica y Computacional,
  Universidad de Granada, E-18071 Granada, Spain}\\
{\large J. Nieves}\\
{\small Instituto de F\'{\i}sica Corpuscular (IFIC), Centro
 Mixto CSIC-Universidad de Valencia} \\{\small Institutos de Investigaci\'on de
 Paterna, Aptd. 22085, E-46071 Valencia, Spain }\\
{\large L. L. Salcedo}\\
 {\small Departamento de F\'{\i}sica At\'omica Molecular y Nuclear, 
  and} \\ {\small
Instituto Carlos I de F{\'\i}sica Te\'orica y Computacional,
  Universidad de Granada, E-18071 Granada, Spain}\\
 {\large L. Tolos} \\
 {\small Institut de Ci\`encies de l'Espai (IEEC/CSIC), Campus Universitat
Aut\`onoma de Barcelona,} \\
{\small Facultat de Ci\`encies, Torre C5, E-08193 Bellaterra, Spain }
    } 
\date{\today}

\maketitle
%
\begin{abstract} 
The anti-$D$ meson self-energy is evaluated self-consistently, using
unitarized coupled-channel theory, by computing the in-medium
meson-baryon $T$-matrix in the $C=-1,S=0$ sector. The heavy
pseudo-scalar and heavy vector mesons, $\bD$ and $\bD^*$, are treated
on equal footing as required by heavy quark spin symmetry. Results for
energy levels and widths of $D^-$ mesic atoms in $^{12}\mbox{C}$,
$^{40}\mbox{Ca}$, $^{118}\mbox{Sn}$ and $^{208}\mbox{Pb}$ are
presented. The spectrum contains states of atomic and of nuclear
types for all nuclei. $\bD^0$-nucleus bound states are also
obtained. We find that, after electromagnetic and nuclear cascade,
these systems end up with the $\bD$ bound in the nucleus, either as a
meson or as part of a exotic $\bD N$ (pentaquark) loosely bound state.
\end{abstract}

\section{Introduction}

Hadronic atoms provide valuable information about in-medium
modification of hadron properties, on hadron-nucleon interaction, and
also on properties of nuclei not easily accessible by other probes, as
the distribution density of neutrons. This field has been the subject
of thorough study, both theoretical and experimental, since long time
ago for pions and anti-kaons \cite{Ericson:1966fm,Friedman:2007zza, 
Nieves:1993ev,GarciaRecio:1991wk,Hirenzaki:2008zz,Gilg:1999qa}, and more
recently for anti-protons
\cite{Wycech:2007jb,Klos:2007is,Trzcinska:2001sy}.  However, for
anti-charmed atoms not much theoretical work exists in the literature.
To our knowledge, only Ref.~\cite{Tsushima:1998ru} studies $D^-$
atoms. There, the $1s$, $2s$ and $1p$ states (neglecting widths) of
$D^-$ in $^{208}$Pb were evaluated using the quark-meson coupling
model of Ref.~\cite{Guichon:1987jp}. $\bD NN$ bound states (rather
than atoms) were predicted in \cite{Yasui:2009bz}. On the experimental
side, the study of anti-$D$ mesic atoms poses a serious challenge. The
study of open charm systems seems timely in view of the forthcoming
experiments by the PANDA \cite{Wiedner:2011mf,PANDA} and CBM
\cite{Aichelin,Staszel:2010zz} Collaborations at the future FAIR
facility at Darmstadt \cite{fair}.

As compared to other mesic atoms, $D^-$ atoms have a number of
specific features that make them worth studying. First, the $\bD$
meson is so heavy that the atomic meson wave function has a sizable
overlap with the nucleus, specially for the low lying levels and for
heavy nuclei. Hence the strong interaction effects are expected to be
larger than for other mesic atoms, even if the optical potentials
themselves are of comparable strength.  Second, $\bD N$ has no lower
hadronic channels for strong interaction decay. This is unlike other
hadron-nucleus bound systems. For instance, in pionic atoms the
channel $\pi NN\to NN$ is available, in $K^-$ atoms $\bar{K} N\to \pi
\Lambda$ and $\pi\Sigma$, in $D^0$-nucleus $D N \to \pi \Lambda_c$ and
$\pi\Sigma_c$, in $\bar{p}$ atoms $\bar{p} N\to$pions, or in
$\eta$-nucleus, $\eta N\to \pi N$.  So, if bound, the $\bD$ remains in the
nucleus until its weak decay is produced. Third, heavy quark spin
symmetry (HQSS), a well established approximate QCD symmetry
\cite{Isgur:1989vq,Neubert:1993mb}, is expected to play an important
role in $D^-$ atoms. One of the consequences of HQSS is that the
$\bar{D}^*$ vector meson degrees of freedom should have some
(important) influence on these systems. Hence, such degrees of freedom
should be incorporated by means of any realistic treatment. This is
automatically achieved in the SU(8) extended Weinberg-Tomozawa model
followed in this work~\cite{GarciaRecio:2008dp,Gamermann:2010zz}. Fourth, all
$t$-channel vector meson exchange models without incorporating HQSS,
that is, not including vector mesons in the coupled-channel space, produce
a featureless real repulsive potential below threshold
\cite{Lutz:2005vx,Tolos:2007vh,JimenezTejero:2011fc}. This scenario is expected to
change when HQSS is enforced.  Indeed, the calculation of
\cite{Yasui:2009bz} identifies an $I=0, J=1/2^-$ $\bD N$ bound state
with $1.4\,\MeV$ of binding energy. The same state is also found in
the SU(8) model of Ref.~\cite{Gamermann:2010zz}. This exotic baryonic state plays
an important role in the $D^-$ atom dynamics. Due to the existence of
this exotic state, the $\bD$ optical potential turns out to be
attractive, dissipative and strongly energy dependent. In addition,
due to the energy dependence, not so relevant in other mesic atoms,
a proper implementation of the electromagnetic interaction, through
minimal coupling, needs to be considered.

The paper is organized as follows. In Sect.~\ref{sec:2} we describe
the calculation  of the $\bD$ self-energy in
nuclear matter
and present our results for the $\bD$ optical potential. We carry out a self-consistent calculation in
symmetric nuclear matter at zero temperature for energies around the
$\bD$ mass. In Sect.~\ref{sec:3} we present our results for the
energies and widths of the $D^-$ mesic atom levels in $^{12}$C,
$^{40}$Ca, $^{118}$Sn and $^{208}$Pb. For this purpose, we solve the
Schr\"odinger equation with a finite nuclei $\bD$ optical potential
obtained from that derived for nuclear matter, in the previous
section, and making use of the local density approximation. In this
section, we also extend our study to the case of $\bD^0$ bound states.
Finally in Sect.~\ref{sec:4}, we discuss possible decay mechanisms of
the bound states, while in Sect.~\ref{sec:5} we summarize the main
conclusions of the present work.

\section{The $\bD$ self-energy and optical potential}
\label{sec:2}

The self-energy in symmetric nuclear matter for the $\bD$ meson is obtained
following a self-consistent procedure in coupled channels, as similarly done
for the $D$ meson \cite{Tolos:2009nn}. The $s$-wave transition potential of the
Bethe-Salpeter equation is derived from an effective Lagrangian that
implements HQSS. This is an approximate QCD symmetry that treats on equal
footing heavy pseudo-scalar and vector mesons
\cite{Isgur:1989vq,Neubert:1993mb}. Therefore, we calculate simultaneously the
self-energy of the $\bD^*$, the HQSS partner of the $\bD$.

As shown in \cite{GarciaRecio:2006wb,GarciaRecio:2005hy}, the
Weinberg-Tomozawa (WT) meson-baryon Lagrangian admits a unique and
natural extension with spin-flavor symmetry for any number of flavors.
In addition to $0^+$ mesons and $1/2^+$ baryons, this requires the
inclusion of $1^-$ mesons and $3/2^+$ baryons. For four flavors this
interaction has SU(8) symmetry and automatically enjoys HQSS in the
$C=-1$ sector. Schematically \cite{GarciaRecio:2008dp,Gamermann:2010zz},
\begin{equation}
{\cal L_{\rm WT}^{\rm SU(8)}} = \frac{1}{f^2} \left((M^\dagger\otimes
M)_{\sesentaytres_a}\otimes (B^\dagger\otimes B)_{\sesentaytres}\right)_{\uno}
.
\label{eq:NOcoupl}
\end{equation}

The tree level amplitudes for different isospin ($I$), total angular momentum
($J$), charm ($C$) and strangeness ($S$) take the form
\begin{equation}
V^{IJSC}_{ab}(\sqrt{s})= D^{IJSC}_{ab}
\frac{2\sqrt{s}-M_a-M_b}{4\,f_a f_b} \sqrt{\frac{E_a+M_a}{2M_a}}
\sqrt{\frac{E_b+M_b}{2M_b}} \,, \label{eq:vsu8break}
\end{equation}
where $M_a$ ($M_b$) and $E_a$ ($E_b$) are, respectively, the mass and the
center of mass energy of the baryon in the $a$ ($b$) channel. The matrix
elements $D^{IJSC}_{ab}$ of the SU(8) WT interaction can be obtained from Wick
contractions using the hadronic wave functions~\cite{GarciaRecio:2008dp} or by means of
the SU(8)$\supset$SU(4)$\otimes$SU(2) Clebsch-Gordan
coefficients~\cite{GarciaRecio:2010vf}. The spin-flavor SU(8) symmetry is strongly broken
in nature and this is incorporated by adopting the physical hadron masses and
different weak decay constants, $f_a$, for non-charmed and charmed,
pseudo-scalar and vector mesons \cite{GarciaRecio:2008dp,Gamermann:2010zz}.

In what follows, we focus in the non-strange ($S=0$) and singly anti-charmed
($C=-1$) sector, where the $\bD N$ and $\bD^* N$ states are embedded. The
channels involved in the coupled-channel calculation are: $\bD N$ and $\bD^*
N$ for $I=0, J=1/2$; $\bD^* N$ for $I=0, J=3/2$; $\bD N$, $\bD^* N$ and $\bD^*
\Delta$ for $I=1, J=1/2$; and $\bD \Delta$, $\bD^* N$ and $\bD^* \Delta$ for
$I=1, J=3/2$.

The amplitudes in nuclear matter [$T^{\rho,IJ}(P_0,{\bm P})$] are obtained by
solving the on-shell Bethe-Salpeter equation with the tree level amplitude
$V^{IJ}(\sqrt{s})$:
\begin{eqnarray}
T^{\rho,IJ}(P) &=& \frac{1}{1-
V^{IJ}(\sqrt{s})\, G^{\rho,IJ}(P)}\,V^{IJ}(\sqrt{s})
,
 \label{eq:scat-rho}
\end{eqnarray}
where the diagonal $G^{\rho,IJ}(P)$ matrix accounts for the
meson-baryon loop in nuclear matter.  The logarithmic divergence in
the vacuum part of the loop function, $G^0(\sqrt{s})$, is removed by
subtraction. Following ~\cite{GarciaRecio:2008dp,Gamermann:2010zz}, we set
$G^{0,IJ}(\sqrt{s}=\mu^{IJ})=0$ with
\begin{equation}
\left (\mu^{IJ}\right)^2 = \alpha
\left(m_{{\rm th}}^2+M^2_{{\rm th}}\right)
.
\label{eq:sp}
\end{equation}
Here $m_{{\rm th}}$ and $M_{{\rm th}}$ are, respectively, the meson and baryon
masses of the hadronic channel with lowest mass threshold for the given $I,J$.
The value of the parameter $\alpha$ is set to one \cite{Hofmann:2005sw}. However, in the
following, we will also vary $\alpha$ to have an estimate of the sensitivity
of our results against changes in the regularization scale.

Nuclear matter effects enter in the meson-baryon loop function
$G^{\rho,IJ}(P)$. One of the sources of density dependence comes from Pauli
blocking. Another source is related to the change of the properties of mesons
and baryons in the intermediate states due to the interaction with nucleons of
the Fermi sea. We proceed as in Ref.~\cite{Tolos:2009nn}, where the most important
changes in matter came from the Pauli blocking of nucleons and from the
self-consistent treatment of the open charm self-energies.

Thus, for the $\bD N$ and $\bD^* N$ channels, the meson-baryon loop function
in matter is given by ~\cite{Tolos:2009nn}:
\begin{eqnarray}
{G^\rho}_{\bD(\bD^*)N}(P)
&=&
G^0_{\bD(\bD^*)N}(\sqrt{s})+
\int \frac{d^3 q}{(2 \pi)^3} \,
\frac{ M_N }{ E_N({\bm p})} \,
\Bigg[ 
\frac{-n({\bm p})}{(P^0 - E_N({\bm p}))^2-\omega({\bm q})^2+i\varepsilon} 
\,  
\label{eq:Glarga} 
\\ &&
+(1-n({\bm p}))
 \Bigg(
\frac{-1/(2 \omega({\bm q}))}
{P^0 -E_N({\bm p})-\omega({\bm q})+i \varepsilon}
+
\int_{0}^{\infty} \, d\omega \,
\frac{S_{\bD(\bD^*)}(\omega,{\bm q})}{P^0 -E_N({\bm p})-\omega+i\varepsilon}
\, \Bigg) \Bigg]  \Bigg|_{{\bm p}={\bm P}-{\bm q}}
, 
\nonumber
\end{eqnarray}
where $E_N({\bm p})=\sqrt{{\bm p}^2+M_N^2}$ is the nucleon energy and
$\omega({\bm q})=\sqrt{{\bm q}^2+m_{\bD(\bD^*)}^2}$ is the
$\bD(\bD^*)$ energy. The free loop function $G^0(\sqrt{s})$ is
corrected in matter by terms proportional to the nucleon Fermi
distribution $n({\bm p})=\Theta(|{\bm p}|-p_F)$ that takes into
account Pauli blocking effects. The quantities ${\bm p}$ and $p_F$ are
the momentum of the nucleon and the Fermi momentum at nuclear density
$\rho$, respectively.   The implementation of
the $\bD$ and $\bD^*$ properties in matter comes through the meson
spectral functions, $S_{\bD(\bD^*)}(\omega,{\bm q})$, which are
defined from the in-medium $\bD$ and $\bD^*$ meson propagators:
\begin{eqnarray}
D^\rho_{\bD (\bD^*)}(q)
&=&
\left ((q^0)^2 -\omega({\bm q})^2-\Pi_{\bD(\bD^*)}(q) \right )^{-1}
,
\nonumber \\
S_{\bD(\bD^*)}(q) &=& -\frac{1}{\pi}\,{\rm Im} D^\rho_{\bD (\bD^*)}(q)
\quad \mbox{(for~$q^0>0$)}
.
\label{eq:Drho}
\end{eqnarray}
The self-energies, $\Pi_{\bD(\bD^*)}(q^0,{\bm q}; \rho)$, are obtained
self-consistently from the in-medium $\bD N$ and $\bD^* N$ effective
interactions as we will show in the following.
As for $\bD \Delta$ and $\bD^* \Delta$ channels, we include the self-energy
of the $\bD$ and $\bD^*$ mesons. Then, the equivalent of
Eq.~(\ref{eq:Glarga}) for those channels reads \cite{Tolos:2009nn}
\begin{eqnarray}
{G^\rho}_{\bD(\bD^*)\Delta}(P) &=&
 G^0_{\bD(\bD^*)\Delta}(\sqrt{s}) 
+\int \frac{d^3 q}{(2 \pi)^3} \,
\frac{ M_\Delta }{ E_\Delta({\bm p})} \,
 \Bigg (
\frac{-1/(2 \omega({\bm q}))}
{P^0 -E_\Delta({\bm p})-\omega({\bm q})+i \varepsilon}
\label{eq:propDD}\\ 
&& 
+
\int_{0}^{\infty} \,
 d\omega \,
\frac{S_{\bD(\bD^*)}(\omega,{\bm q})}{P^0 -E_\Delta({\bm p})
-\omega+i\varepsilon}
\, \Bigg )  \Bigg| _{{\bm p}={\bm P}-{\bm q}}
,
\nonumber
\end{eqnarray}
with $E_\Delta({\bm p})=\sqrt{{\bm p}^2+M_\Delta^2}$. The effect of
the vacuum width of the $\Delta$ has not been included. The strong
width of the $\bD^*$ is very small, as a consequence of HQSS.

The $\bD$ self-energy in symmetric nuclear matter is  obtained by summing
the different isospin transition amplitudes for $\bD N$ over the nucleon Fermi
distribution as
\begin{eqnarray}
\Pi_{\bD}(q^0,{\bm q}; \rho) &=& \int_{p \leq p_F} 
\frac{d^3p}{(2\pi)^3} \,
   \Big[\, T^{\rho,0,1/2}_{\bD N} (P^0,{\bm P}) +
3 \, T^{\rho,1,1/2}_{\bD N}(P^0,{\bm P}) \Big]
.
\label{eq:selfd}
\end{eqnarray}
Simultaneously, the $\bD^*$ meson self-energy is derived from the sum
over the $\bD^*N$ amplitudes as\footnote{We neglect the enhancement in
the $\bD^*$ width due to coupling to $\bD\pi$ (and their medium
corrections). The analogous mechanism for $\bar{K}^*\to\bar K \pi$ 
was considered in \cite{Tolos:2010fq}.}
\begin{eqnarray}
\Pi_{\bD^*}(q^0,{\bm q}; \rho\,) &=& \int _{p \leq p_F} \frac{d^3p}{(2\pi)^3} \,
\Bigg[ \frac{1}{3} \, T^{\rho,0,1/2}_{\bD^* N}(P^0,{\bm P}) +
T^{\rho,1,1/2}_{\bD^* N}(P^0,{\bm P}) 
\nonumber \\
&& 
+   \frac{2}{3} \,
T^{\rho,0,3/2}_{\bD^* N}(P^0,{\bm P}) + 
2 \, T^{\rho,1,3/2}_{\bD^* N}(P^0,{\bm P})\Bigg]
.
\label{eq:selfds}
\end{eqnarray}
\noindent
In the above equations, $P^0=q^0+E_N({\bm p})$ and ${\bm
P}={\bm q}+{\bm p}$ are the total energy and momentum of the
meson-nucleon pair in the nuclear matter rest frame, and $(q^0,{\bm
q})$ and $(E_N,{\bm p})$ stand for the energy and momentum of the
meson and nucleon, respectively, in that frame.
  As mentioned
previously, those self-energies are determined self-consistently since they
are obtained from the in-medium amplitudes which contain the meson-baryon loop
functions, and those quantities themselves are functions of the self-energies.

\begin{figure}[t]
\begin{center}
\includegraphics[width=1.0\textwidth]{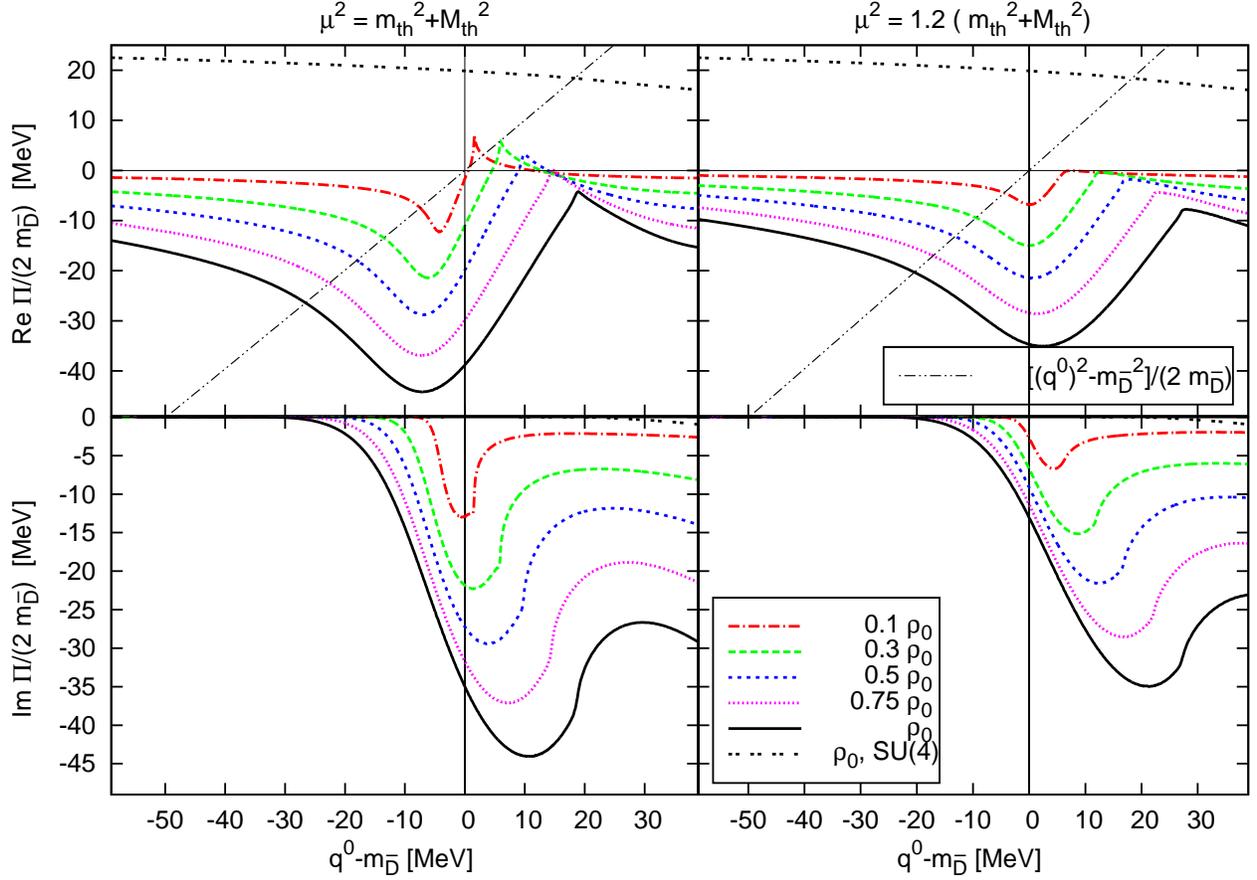}
\caption{\small Real and imaginary parts of the $\bD$ self-energy over
$2m_\bD$, at ${\bm q}=0$, as functions of the meson energy $q^0$ for
different densities and two subtraction points with $\alpha=1$ (left
panels) and $\alpha=1.2$ (right panels). The oblique line is the
function $((q^0)^2-m_\bD^2)/(2 m_\bD)$. The  SU(4)  $\bD$ self-energy
obtained in Ref.~\cite{Tolos:2007vh} for normal nuclear matter
density is also displayed.}
\label{fig:self}
\end{center}
\end{figure}

We are interested in studying possible $\bD$ bound states in nuclei.
Therefore, we concentrate on the self-energy for $q^0$ around the $\bD$
mass. In Fig.~\ref{fig:self} we show the $\bD$ self-energy over $2m_\bD$, as a
function of the $\bD$ energy, for various nuclear densities $\rho$, and with
the $\bD$ meson momentum $\bm q=0$. We display results for two values for the
subtraction point (see Eq.~(\ref{eq:sp})): $\alpha=1$ (left panels) and
$\alpha=1.2$ (right panels). For comparison, we also show results for the
SU(4) WT model of Ref.~\cite{Tolos:2007vh} at normal nuclear density,
$\rho_0=0.17 \ {\rm fm^{-3}}$.

It is worth noticing a resonant structure (more pronounced for the
preferred value $\alpha=1$) close to the $\bD N$ threshold, which will
be of up-most importance for the study of $\bD$ bound states. This
structure results from a pole in the free space amplitude of the
sector $I=0,J=1/2$ at $2805\,{\rm MeV}$ (a weakly bound pentaquark
state) that strongly couples to $\bD N$ and $\bD^* N$ states
\cite{Gamermann:2010zz} (also found in \cite{Yasui:2009bz}). For reference
we will call this state $X(2805)$\footnote{This state is bound by only
about 1 MeV in the free space, and it is one of the most interesting
predictions of Ref.~\cite{Gamermann:2010zz}. Moreover, it appears as a
consequence of considering heavy vector meson degrees of freedom, as
required by HQSS. Indeed, the diagonal $\bD N$ WT interaction is zero
in this sector and thus, the $X(2805)$ is  generated thanks to the
coupled channel dynamics between the $\bD N$ and $\bD^* N$
pairs. Thus, this bound state is absent in the free space SU(4) WT
model of Ref.~\cite{Hofmann:2005sw} in which is based the nuclear
medium approach of Ref.~\cite{Tolos:2007vh}.}. The situation
has some similarities with the $\bar{K}N$ interaction, which is
governed by the $\Lambda(1405)$ resonance. The $\Lambda(1405)$
dominates the behavior of the $\bar K N$ interaction close to
threshold similarly to the pole in $2805\,{\rm MeV}$ for the $\bar D
N$ amplitude. However, the $\Lambda(1405)$ can decay into $\pi
\Sigma$, whereas the $X(2805)$ is below all thresholds for strong
interaction decay. The exotic $X(2805)$ has a HQSS partner with
$I=0,J=3/2$, a $\bD^* N$ bound state with mass $2922\,\MeV$, as seen
in Ref.~\cite{Gamermann:2010zz}.

In contrast to the SU(8) scheme and as mentioned above, a resonant
structure is not observed in the SU(4) WT model of
Ref.~\cite{Tolos:2007vh}. The SU(4) amplitude is repulsive
and shows a smooth behavior as a function of the energy. A similar
repulsive effect was observed in the $t$-channel vector meson exchange
models of Refs. \cite{Lutz:2005vx,JimenezTejero:2011fc}.

Due to the strong energy dependence of the in-medium effective interaction in
the SU(8) WT scheme close to threshold, any slight change in the parameters of
the model as well as in the self-consistent procedure may have strong
consequences on the formation of $\bD$-nucleus bound states. In order to mimic
those changes, we have slightly varied the subtraction point, namely, to
$\alpha=1.2$. In this way we study two very distinct situations for the
formation of bound states and set our theoretical uncertainties.

The $\bD$ self-energy is evaluated in infinite nuclear matter. In
finite nuclei we use the local density approximation (LDA),
substituting $\rho$ by $\rho (r)$, which is the local density at each
point in the nucleus. For the $s$-wave, as it is the case here, it was
shown in Ref.~\cite{Nieves:1993ev} that the LDA gave the same results as a
direct finite nucleus calculation. The LDA $\bD$ self-energy allows to
define a local optical potential. In mesic atoms this optical
potential is often taken to be energy independent and fixed to its value at
threshold ($q^0=m_\bD,\, {\bm q}=0$). However, both the real and the
imaginary parts of the $\bD$ self-energy, around the $\bD$-meson mass,
show a pronounced energy dependence, as can be appreciated in
Fig.~\ref{fig:self}.  Hence, a realistic determination of the $\bD$
bound states should take this energy dependence into account, as done
previously for $\eta$- and $D^0$-nucleus systems
\cite{GarciaRecio:2002cu,GarciaRecio:2010vt}.  Thus, we use an energy dependent
optical potential defined as:
\begin{equation}
  V_{\rm opt}(r,q^0) = \frac{1}{2 q^0} \Pi_{\bD}(q^0,{\bm q}=0,\,\rho(r))
.
\label{eq:UdepE}
\end{equation}

Most of the imaginary part for $q^0<m_\bD$ displayed in
Fig.~\ref{fig:self} comes from particle-hole production and this is
allowed due to the attractive potential felt by the $\bD$ in the
medium.  The quantity $((q^0)^2-m_\bD^2)/(2 m_\bD)$ is displayed in
Fig.~\ref{fig:self} by a dashed-dotted-dotted solid line.  The
leftmost crossing point of this line with the real part of the
self-energy (divided by $2m_\bD$) signals the opening of the
$\bD$-particle-hole threshold.  For the energies displayed,
$((q^0)^2-m_\bD^2)/(2 m_\bD)$ is essentially $q^0-m_\bD=E$, the non
relativistic energy of the $\bD$ (and so almost a straight
line). Therefore, the difference $E-{\rm Re}\, V_{\rm opt}(E)$
corresponds to the kinetic energy of the non-relativistic problem. The
two lines $E$ and ${\rm Re}\,V_{\rm opt}(E)$ cross at the classical
turning point.  Roughly speaking, there should not be imaginary part
in the classically forbidden region $E<{\rm Re}\, V_{\rm opt}(E)$, as
there is no available phase space for decay (i.e., no kinetic energy
to expend). Also, the bound states should appear predominantly for
energies fulfilling the condition $E>{\rm Re}\, V_{\rm opt}(E)$, since
the expectation value of the kinetic energy in the bound state cannot
be negative. Of course, these arguments are only qualitative because
the optical potential is complex and strongly energy dependent. This
allows for the $\bD$ in the medium to have some non intuitive
behavior. For instance, a $\bD$ with energy $E$ can eject a
particle-hole going to a lower energy $E^\prime$, and yet end up with
more kinetic energy to expend, provided $E-E^\prime < {\rm Re}\,
V_{\rm opt}(E) - {\rm Re}\, V_{\rm opt}(E^\prime)$.

Due to the $\bD N$ bound state close to threshold, $X(2805)$, the low
density approximation $ T \rho$ breaks down very early. For a given
value of the energy the density dependence of the optical potential is
far from linear.  For subsequent use, we have computed the optical
potential for several densities (those in Fig.~\ref{fig:self}) and a
fine lattice of energies, and have used an interpolation procedure for
other values of density and energy. The presence of the bound
state/resonance prevented the self-consistent procedure to be
continued for densities below $0.1\,\rho_0$.

\section{Results}
\label{sec:3}

\begin{table}[t]
\caption{$D^-$-atom binding energies $B$ [keV], calculated using only
the Coulomb potential for several nuclei and orbital angular momenta
($L$). The Coulomb degeneracy is corrected by nuclear finite size and
vacuum polarization effects.}
\label{tab:coul}
\footnotesize
\vspace*{.4cm}
{
\begin{tabular}{|c|rrrrr|}
\hline \jtstrut
  $L$  &\multicolumn{5}{|c|}{$^{12}$C}\\
\hline \jtstrut
 0 & 1158 & 333 & 155 & 89 & 58 \\
 1 &  384 & 171 &  96 & 61 &    \\
 2 &  171 &  96 &  62 &    &    \\
 3 &   96 &  62 &     &    &    \\
 4 &   62 &     &     &    &    \\
\hline
\end{tabular}
\\
\begin{tabular}{|c|rrrrrrrrr|}
\hline \jtstrut
  $L$  &  \multicolumn{9}{|c|}{$^{40}$Ca} \\
\hline \jtstrut
 0 & 6073 & 2626 & 1412 & 875 & 594 & 429 & 324 & 254 & 204 \\
 1 & 3664 & 1768 & 1038 & 682 & 482 & 359 & 277 & 220 &     \\
 2 & 2052 & 1156 &  742 & 517 & 381 & 292 & 231 &     &     \\ 
 3 & 1190 &  760 &  528 & 388 & 297 & 234 &     &     &     \\
 4 &  762 &  529 &  388 & 297 & 235 &     &     &     &     \\
 5 &  529 &  388 &  297 & 235 &     &     &     &     &     \\
\hline
\end{tabular}
\\
\begin{tabular}{|c|rrrrrrrrr|}
\hline \jtstrut
  $L$ & \multicolumn{9}{|c|}{$^{118}$Sn}\\
\hline \jtstrut
 0 & 14536 & 9260 & 5919 & 4038 & 2912 & 2194 & 1710 & 1369 & 1121 \\
 1 & 11665 & 7230 & 4766 & 3349 & 2475 & 1901 & 1505 & 1220 &      \\
 2 &  8996 & 5611 & 3821 & 2764 & 2091 & 1637 & 1315 &      &      \\
 3 &  6650 & 4321 & 3051 & 2272 & 1758 & 1401 &      &      &      \\
 4 &  4760 & 3290 & 2429 & 1855 & 1469 &      &      &      &      \\
 5 &  3404 & 2494 & 1907 & 1506 &      &      &      &      &      \\ 
 6 &  2509 & 1920 & 1516 &      &      &      &      &      &      \\
 7 &  1920 & 1517 &      &      &      &      &      &      &      \\
\hline
\end{tabular}
\\
\begin{tabular}{|c|rrrrrrrrr|}
\hline \jtstrut
  $L$ 
                              & \multicolumn{9}{|c|}{$^{208}$Pb}\\
\hline \jtstrut
 0 & 21485 & 16006 & 11454 & 8371 & 6317 & 4915 & 3923 & 3200 & 2658 \\
 1 & 18680 & 13502 &  9682 & 7180 & 5503 & 4341 & 3506 & 2888 &      \\
 2 & 15929 & 11256 &  8167 & 6151 & 4787 & 3825 & 3124 & 2599 &      \\
 3 & 13276 &  9310 &  6863 & 5258 & 4154 & 3362 & 2777 &      &      \\
 4 & 10787 &  7654 &  5749 & 4482 & 3595 & 2948 &      &      &      \\
 5 &  8550 &  6249 &  4801 & 3813 & 3104 & 2578 &      &      &      \\
 6 &  6662 &  5063 &  3991 & 3232 & 2672 &      &      &      &      \\
 7 &  5190 &  4089 &  3306 & 2729 &      &      &      &      &      \\
\hline
\end{tabular}\\
}\normalsize
\end{table}

\begin{table}[t]
\caption{Complex $D^-$-atom energies $(B,\Gamma/2)$ for several
  nuclei.  The calculation includes Coulomb potential plus the SU(8)
  energy dependent optical potential with $\alpha=1$ and a nucleon gap
  of $8\,\MeV$ (see Fig.~\ref{fig:self}).  To preserve the structure of
  the atomic states, the states of nuclear type (when there exist) are
  displayed in the first line that corresponds to each of the angular
  momentum entries. The second line in the entry contains the energies
  of the atomic states.  Results of Ref. \cite{Tsushima:1998ru} for
  the binding energies in ${}^{208}$Pb are also shown for model $\tilde{V}^q_\omega$ between
  square brackets and for model $V^q_\omega$ without brackets. }
\label{tab:a18}
\footnotesize
\vspace*{.4cm}

\begin{tabular}{|c|cccc|}
\hline 
\multicolumn{5}{|c|}
{$^{12}$C,   $\alpha=1.0$, ${\rm gap} = 8\,\MeV$ }
\cr
\hline \jtstrut
 $L$ & 
\multicolumn{4}{|c|}{$(B,\Gamma/2)$ [keV] }\cr
\hline \jtstrut
 0 &
     $( 22800,    199 )$  & & &
    \cr
    &
     $(        585,      51 )$   &
     $(        222,      12 )$   &
     $(        116,       5 )$   &
     $(         71,       2 )$   \cr
 1 & 
   $(16000,   1100 )$ & & &
    \cr
    &
   $(    353,      9 )$  &  
   $(    160,      3 )$  &  
   $(     91,      1 )$  &  
   $(     59,      1 )$  \cr 
2 &
   $(    171,      0.1 )$&
   $(     96,      0.1 )$&
   $(     62,      0.0 )$&\cr
3 &
   $(     96,      0 )$ &
   $(     62,      0 )$ & &\cr
4 &
   $(     62,      0 )$ & & &\cr
\hline
\end{tabular}\\

\begin{tabular}{|c|c c c c c c c c |}
\hline 
\multicolumn{9}{|c|}
{$^{40}$Ca, $\alpha=1.0$, ${\rm gap} = 8\,\MeV$ }\cr
\hline \jtstrut
$L$  & 
\multicolumn{8}{|c|}{$(B,\Gamma/2)$ [keV] }\cr
\hline \jtstrut
 0 &
 $( 32177,  48)$ &
 $( 22764, 978)$ & & & & & &
 \cr
   &
 $(     2556,      184)$ & 
 $(     1305,       85)$ & 
 $(      806,       45)$ & 
 $(      550,       27)$ & 
 $(      400,       17)$ & 
 $(      304,       12)$ & 
 $(      239,        8)$ &
 $(      193,        6)$
 \cr
 1 &
 $(    29486,      306)$ & 
 $(    17022,     2956)$ & &&&&&
 \cr
   &
 $(     2244,      138)$ & 
 $(     1187,       66)$ & 
 $(      748,       36)$ & 
 $(      517,       22)$ & 
 $(      379,       14)$ & 
 $(      291,       10)$ & 
 $(      230,        7)$ & 
 $(      186,        5)$  
 \cr
 2 &
 $(    25282,      892)$ &&&&&&&
 \cr
   &
 $(     1716,       67)$ & 
 $(      981,       36)$ & 
 $(      643,       21)$ & 
 $(      456,       13)$ & 
 $(      341,        9)$ & 
 $(      264,        6)$ & 
 $(      211,        5)$ & 
 $(      172,        3)$ 
 \cr
 3 &
  $(   18596,   1952)$ &&&&&&&
 \cr
   &
 $(     1160,       13)$ & 
 $(      738,       10)$ & 
 $(      512,        7)$ & 
 $(      377,        5)$ & 
 $(      289,        3)$ & 
 $(      229,        2)$ & 
 $(      185,        2)$ & 
 \cr
 4 &
 $(      761,        1)$ & 
 $(      528,        1)$ & 
 $(      388,        1)$ & 
 $(      297,        0)$ & 
 $(      234,        0)$ & 
 $(      190,        0)$ & & 
 \cr
 5 &
 $(      528,        0)$ & 
 $(      388,        0)$ & 
 $(      297,        0)$ & 
 $(      235,        0)$ & 
 $(      190,        0)$ & & &
 \cr 
 \hline
\end{tabular}\\
%

\begin{tabular}{|c|c c c c c c c c |}
\hline 
\multicolumn{9}{|c|}
{$^{118}$Sn,   $\alpha=1.0$, ${\rm gap} = 8\,\MeV$}\cr
\hline \jtstrut
 $L$ & 
\multicolumn{8}{|c|}{$(B,\Gamma/2)$ [keV] }\cr
\hline \jtstrut
 0 &
  $(   40768,       25)$ &
  $(   35777,      271)$ & 
  $(   28200,     1527)$ & & & & &
\cr
   &
  $(    6102,      414)$ & 
  $(    3727,      286)$ & 
  $(    2619,      206)$ & 
  $(    2004,      134)$ & 
  $(    1576,       92)$ & 
  $(    1272,       67)$ & 
  $(    1048,       51)$ &
  $(     878,       39)$ 
\cr
 1 &
  $(   38227,       69)$ &
  $(   32794,      651)$ &
  $(   23353,     4033)$ & &&&&
\cr
   &
  $(    5853,      389)$ & 
  $(    3612,      268)$ & 
  $(    2558,      192)$ & 
  $(    1960,      125)$ & 
  $(    1544,       87)$ & 
  $(    1248,       64)$ & 
  $(    1030,       48)$ & 
  $(     864,       37)$ 
\cr
 2  &
  $(   28504,     1236)$ &&&&&&&
\cr
  &
  $(    5378,      337)$ & 
  $(    3390,      234)$ & 
  $(    2439,      165)$ & 
  $(    1873,      109)$ & 
  $(    1481,       77)$ & 
  $(    1201,       57)$ & 
  $(     994,       43)$ & 
  $(     836,       33)$ 
\cr
 3 &
  $(    4726,      260)$ & 
  $(    3081,      187)$ & 
  $(    2262,      129)$ & 
  $(    1747,       88)$ & 
  $(    1390,       63)$ & 
  $(    1134,       47)$ & 
  $(     943,       36)$ & 
  $(     796,       28)$   
\cr
 4 &
  $(    3968,      168)$ & 
  $(    2709,      130)$ & 
  $(    2033,       89)$ & 
  $(    1587,       63)$ & 
  $(    1275,       46)$ & 
  $(    1048,       35)$ & 
  $(     877,       27)$ &
  $(     745,       21)$   
\cr
 5 &
  $(    3189,       80)$ & 
  $(    2308,       68)$ & 
  $(    1769,       50)$ & 
  $(    1403,       37)$ & 
  $(    1141,       28)$ & 
  $(     947,       22)$ & 
  $(     799,       17)$ &  
\cr
 6 &
  $(    2482,       20)$ & 
  $(    1889,       21)$ & 
  $(    1489,       18)$ & 
  $(    1205,       15)$ & 
  $(     995,       12)$ & 
  $(     836,       10)$ & &
 
\cr
 7 &
  $(    1919,        2)$ & 
  $(    1515,        2)$ & 
  $(    1226,        3)$ & 
  $(    1013,        2)$ & 
  $(     850,        2)$ & & &
\cr
\hline
\end{tabular}\\

%
\begin{tabular}{|c|cccccc|cc|}
\hline 
\multicolumn{7}{|c|}
{$^{208}$Pb,   $\alpha=1.0$, ${\rm gap} = 8\,\MeV$ }
& 
\multicolumn{2}{|c|}{$^{208}$Pb, ~~Ref.~\cite{Tsushima:1998ru} }
 \cr
\hline \jtstrut
 $L$ & 
\multicolumn{6}{|c|}{$(B,~\Gamma/2)$ [keV] }
&
\multicolumn{2}{|c|}{$B$ [keV] }
\cr
\hline \jtstrut 
0 &
 $( 47203,    21 )$ &
 $( 42781,   121 )$ & 
 $( 37644,   636 )$ &
 $( 30343,  3755 )$ & 
& & $35.2\times 10^3$ & $30\times 10^3$ 
 \cr
 &
 $(  9418,   606 )$ & 
 $(  6375,   496 )$ & 
 $(  4912,   334 )$ & 
 $(  3892,   243 )$ & 
 $(  3160,   184 )$ & 
 $(  2618,   142 )$ 
 & $[10.6\times 10^3]$ & $[7.7 \times 10^3]$
 \cr
1 & 
 $( 45065,    45 )$ &
 $( 40385,   265 )$ & 
 $( 34634,  1370 )$ & & &
 & $32.1\times 10^3$ &
 \cr
 &
 $(  9206,   590 )$ &
 $(  6283,   474 )$ & 
 $(  4841,   322 )$ & 
 $(  3839,   235 )$ & 
 $(  3120,   178 )$ & 
 $(  2587,   138 )$ 
& $[10.2\times 10^3]$ &
 \cr
2 &
 $( 37845,   545 )$ &
 $( 30879,  3098 )$ & & & &
 & &
 \cr
 &
 $(  8793,   557 )$ &
 $(  6096,   434 )$ &
 $(  4700,   299 )$ &
 $(  3734,   219 )$ &
 $(  3040,   167 )$ &
 $(  2526,   130 )$
& &
 \cr
3 &
 $(  8204,   504 )$ &
 $(  5813,   379 )$ &
 $(  4491,   266 )$ &
 $(  3579,   197 )$ &
 $(  2924,   150 )$ &
 $(  2436,   117 )$
& &
 \cr
4 &
 $(  7480,   434 )$ &
 $(  5437,   314 )$ &
 $(  4219,   225 )$ &
 $(  3379,   168 )$ &
 $(  2773,   129 )$ &
 $(  2319,   102 )$ 
& &
 \cr
5 &
 $(  6676,   344 )$ &
 $(  4977,   242 )$ &
 $(  3891,   177 )$ &
 $(  3139,   135 )$ &
 $(  2592,   105 )$ &
 $(  2180,    94 )$
& &
 \cr
6 &
 $(  5868,   214 )$ &
 $(  4443,   165 )$ &
 $(  3518,   127 )$ &
 $(  2867,    99 )$ &
 $(  2387,    79 )$ &
 $(  2020,    63 )$
& &
 \cr 
7 &
 $(  4964,   101 )$ &
 $(  3865,    92 )$ &
 $(  3115,    76 )$ &
 $(  2571,    62 )$ &
 $(  2162,    51 )$ &
& &
 \cr
\hline
\end{tabular}\\

\end{table}

\begin{table}[t]
\caption{Binding energy and half-widths, $(B,\Gamma/2)$, in keV, of
  $\bD^0$-nucleus levels for $^{12}$C, $^{40}$Ca, $^{112}$Sn and
  $^{208}$Pb and several angular momenta. The bound levels have been
  obtained with the SU(8) energy dependent optical potential of
  Fig.~\ref{fig:self} with $\alpha=1$ and a nucleon gap of $8\,\MeV$.
  Results of Ref. \cite{Tsushima:1998ru} for ${}^{208}$Pb are also shown for
  the model $V^q_\omega$ (the model $\tilde{V}^q_\omega$ does not give
  rise to any bound states). }
\label{tab:nocoul}
\footnotesize
\vspace*{.4cm}

\begin{tabular}{|c|c |}
\hline 
\multicolumn{2}{|c|}
{$^{12}$C, $\bD^0$,  $\alpha=1.0$, ${\rm gap} = 8\,\MeV$
 \phantom{$\hat{\dot{1}}$}
}
\cr
\hline \jtstrut
 $L$ & 
\multicolumn{1}{|c|}{$(B,\Gamma/2)$ [keV] }\cr
\hline \jtstrut
 0 &
  $(   18779 ,    176 )$ 
\cr
 1 &
  $( 12324 ,    918 )$ 
\cr
\hline
\end{tabular}\\

\begin{tabular}{|c| c c |}
\hline 
\multicolumn{3}{|c|}
{$^{40}$Ca, $\bD^0$,  $\alpha=1.0$, ${\rm gap} = 8\,\MeV$
 \phantom{$\hat{\dot{1}}$}
}\cr
\hline \jtstrut
 $L$ & 
\multicolumn{2}{|c|}{$(B,\Gamma/2)$ [keV] }\cr
\hline \jtstrut
 0 &
  $( 22549 ,     38 )$ &
  $( 14769 ,    874  )$
\cr
 1 &
  $(  20158 ,    241 )$ &
  $(   8971 ,   1957 )$ 
\cr
 2  &
  $(  16457,    679 )$ &
\cr
 3  &
  $(  10370 ,   1480 )$ &
\cr
\hline
\end{tabular}\\

\begin{tabular}{|c|c c c |}
\hline 
\multicolumn{4}{|c|}
{$^{118}$Sn, $\bD^0$,  $\alpha=1.0$, ${\rm gap} = 8\,\MeV$
 \phantom{$\hat{\dot{1}}$}
}\cr
\hline \jtstrut
 $L$ & 
\multicolumn{3}{|c|}{$(B,\Gamma/2)$ [keV] }\cr
\hline \jtstrut
 0 &
  $(  23339 ,     17 )$ &
  $(  18590 ,    143 )$ & 
  $(  12510 ,   907 )$ 
\cr
 1 &
 &
  $(  15882 ,    355 )$ &
  $(   8939 ,   2004 )$ 
\cr
 2  &
  $(  19317,    106 )$ &
  $(  13009 ,    787 )$ &
\cr
 3  &
  $(  16941 ,    241 )$ & 
  $(   9770 ,   1677 )$ & 
\cr
\hline
\end{tabular}\\

\begin{tabular}{|c|cccc|cc|}
\hline 
\multicolumn{5}{|c|}
{$^{208}$Pb, $\bD^0$,  $\alpha=1.0$, ${\rm gap} = 8\,\MeV$
 \phantom{$\hat{\dot{1}}$}
} 
&
\multicolumn{2}{|c|}{$^{208}$Pb, ~~Ref.~\cite{Tsushima:1998ru} }
\cr
\hline \jtstrut
 $L$ & 
\multicolumn{4}{|c|}{$(B,\Gamma/2)$ [keV] }
&
\multicolumn{2}{|c|}{$B$ [keV] }
\cr
\hline \jtstrut
 0 &
  $(  23381 ,     13 )$ &
  $(  20085 ,     66 )$ & 
  $(  15569 ,    358 )$ &
  $( 9951 ,   1633 ) $  &
  $25.4\times 10^3$ & $19.7\times 10^3$ 
\cr
 1 &
  &
  $( 18102 ,    146 )$ &
  $(  13086 ,   747 )$ &
  $(   5823 ,  3675 )$ &
  $23.1\times 10^3$ & 
\cr
 2  &
  $(  20671 ,     50 )$ &
  $(  16002 ,    304 )$ &
  $(  10336 ,   1494 )$ &
& &
\cr
 3  &
  $(   19006 ,    100 )$ & 
  $(   13770 ,    608 )$ & 
  $(    6682 ,   3208 )$ & 
& & 
\cr
\hline
\end{tabular}\\

\end{table}

\begin{table}[t]
\caption{Same 
as in Table \ref{tab:a18}, but only for  $^{12}$C and different
combinations of  $\alpha$ and the nucleon gap. }
\label{tab:Cx}
\footnotesize
\vspace*{.4cm}
\begin{tabular}{|c|c c c c |}
\hline \jtstrut
  & 
\multicolumn{4}{|c|}
{$^{12}$C, ~  $\alpha=1.2$, ~ ${\rm gap} = 8\,\MeV$ }\cr
\hline \jtstrut
 $L$ & 
\multicolumn{4}{|c|}{$(B,\Gamma/2)$ [keV] }\cr
\hline \jtstrut
 0 &
     $( 16300 ,    104 )$  &
     $(       1829 ,     949 )$   & & 
    \cr
    &
     $(        373 ,      90 )$   &
     $(        167 ,      28 )$   &
     $(         94 ,      12 )$   &
     $(         61 ,       6 )$   \cr
 1 & 
   $(  9300 ,  595 )$  &   & &
    \cr
    &
   $(    354 ,     11 )$  &  
   $(    160 ,      4 )$  &  
   $(     91 ,      2 )$  &  
   $(     59 ,      1 )$  \cr 
2 &
   $(    171 ,      1 )$&
   $(     96 ,      0.6 )$&
   $(     62 ,      0.3 )$&\cr
3 &
   $(     96 ,      0 )$ &
   $(     62 ,      0 )$ & &\cr
4 &
   $(     62 ,      0 )$ & & &\cr
\hline
\end{tabular}\\
\begin{tabular}{|c|c c c c |}
\hline \jtstrut
  & 
\multicolumn{4}{|c|}
{$^{12}$C, ~  $\alpha=1.0$, ~${\rm gap} = 0\,\MeV$ }\cr
\hline \jtstrut
 $L$ & 
\multicolumn{4}{|c|}{$(B,\Gamma/2)$ [keV] }\cr
\hline \jtstrut
 0 &
     $( 22700 ,   2100)$  & & &
   \cr
   & 
     $(        555 ,      47 )$   &
     $(        215 ,      12 )$   &
     $(        113 ,       5 )$   &
     $(         70 ,       2 )$   \cr
 1 & 
   $( 15700 ,   6900 )$  & & &
  \cr
  &
   $(    348 ,     11 )$  &  
   $(    158 ,      4 )$  &  
   $(     91 ,      2 )$  &  
   $(     59 ,      1 )$  \cr 
2 &
   $(    171 ,      0.2 )$&
   $(     96 ,      0.1 )$&
   $(     62 ,      0.1 )$&\cr
3 &
   $(     96 ,      0 )$ &
   $(     62 ,      0 )$ & &\cr
4 &
   $(     62 ,      0 )$ & & &\cr
\hline
\end{tabular}\\
\end{table}

\begin{table}[t]
\caption{ $D^-$-atom binding energies $B$ in keV, calculated using the
SU(4) optical potential of Ref.~\cite{Tolos:2007vh} together
with  the Coulomb interaction.}
\label{tab:su4}
\footnotesize
\vspace*{.4cm} {
\begin{tabular}{|c|rrrr|}
\hline \jtstrut
  $L$  &\multicolumn{4}{|c|}{$^{12}$C}\\
\hline \jtstrut
 0 & 666 & 241 & 123 & 75 \\
 1 &    364 & 164 & 93 & \\
 2 & 171 & 96 &         &     \\
 3 & 96 &         &     &     \\
\hline
\end{tabular}
\\
\begin{tabular}{|c|rrrrr|}
\hline \jtstrut
  $L$  &
                        \multicolumn{5}{|c|}{$^{40}$Ca}\\
\hline \jtstrut
 0 & 
    2946 & 1487 & 903 & 607 & 436 \\
 1 &
    2505 & 1315 & 818 & 559 & 407 \\
 2 &                              
    1816 & 1038 & 677 & 477 & \\
 3 &
    1173 & 748  & 519 & & \\
 4 &
     762 & 528  & & & \\
 5 &
     529 & & & & \\ 
\hline
\end{tabular}
\\
\begin{tabular}{|c|rrrrrrr|}
\hline \jtstrut
  $L$ & \multicolumn{7}{|c|}{$^{118}$Sn}\\
\hline \jtstrut
 0 & 7024 & 4442 & 3106 & 2303 & 1778 & 1415 & 1153 \\
 1 & 6692 & 4255 & 2988 & 2224 & 1722 & 1374 & 1122 \\
 2 & 6068 & 3911 & 2775 & 2082 & 1623 & 1302 & 1069 \\
 3 & 5219 & 3453 & 2493 & 1895 & 1492 & 1207 & \\
 4 & 4250 & 2929 & 2168 & 1677 & 1338 & & \\
 5 & 3295 & 2394 & 1828 & 1444 & & & \\
 6 & 2498 & 1905 & 1502 & & & & \\
 7 & 1920 & 1516 & & & & & \\
\hline
\end{tabular}
\\
\begin{tabular}{|c|rrrrrrrrr|}
\hline \jtstrut
  $L$ & \multicolumn{9}{|c|}{$^{208}$Pb}\\
\hline \jtstrut
 0 & 10787 & 7503 & 5662 & 4612 & 4060 & 3404 & 2826 & 2374 & 2021 \\
 1 & 10520 & 7318 & 5491 & 4305 & 3481 & 2881 & 2430 & 2080 & \\
 2 & 10002 & 6980 & 5246 & 4113 & 3320 & 2740 & 2302 & & \\
 3 &  9261 & 6513 & 4922 & 3876 & 3140 & 2600 & 2190 & & \\
 4 &  8333 & 5943 & 4535 & 3598 & 2933 & 2440 & 2064 & & \\
 5 &  7272 & 5301 & 4101 & 3287 & 2701 & 2263 & & & \\
 6 &  6146 & 4619 & 3640 & 2955 & 2453 & 2072 & & & \\
 7 &  5051 & 3937 & 3170 & 2614 & 2196 & & & & \\
\hline
\end{tabular}
\\
}\normalsize
\end{table}

We look first for $D^-$-nucleus bound states by solving the
Schr\"odinger equation:
 \begin{equation}
   \left[ -\frac{{\bm \nabla}^2}{2 m_{\rm red}} +  V_{\rm{coul}}(r)
+  V_{\rm{opt}}(r) \right] \Psi \,
  = (-B-i \Gamma /2) \Psi
.
\label{eq:SchE}
 \end{equation}

In this equation, $B$ is the binding energy ($B>0$), $\Gamma$ the width
of the bound state and $m_{\rm red}$ is the $\bD$-nucleus reduced
mass. $V_{\rm coul}(r)$ is the Coulomb potential of the $D^-$
including the nucleus finite size and the Uehling vacuum
polarization. $V_{\rm{opt}}(r)$ is the energy dependent optical
potential. Because the electromagnetic interaction is introduced by means of
the minimal coupling prescription (to be consistent with gauge invariance and
electric charge conservation), $V_{\rm coul}(r)$ must be introduced wherever
the energy is present. So the energy dependent optical potential of
Eq.~(\ref{eq:UdepE}) is applied with argument $q^0=m_\bD-B-V_{\rm coul}(r)$.

The non relativistic approximation is used since the $\bD$-meson optical
potential is much smaller than its mass, and we expect the relativistic
corrections to be tiny and certainly smaller than the theoretical
uncertainties of the interaction. In the same approximation the denominator
$2q^0$ in Eq.~(\ref{eq:UdepE}) can also be set to $2m_\bD$.

We solve the Schr\"odinger equation in coordinate space by using a
numerical algorithm~\cite{Oset:1985tb,GarciaRecio:1989xa}, which has
been extensively tested in similar problems of
pionic~\cite{Nieves:1993ev,GarciaRecio:1991wk} and
anti-kaonic~\cite{Baca:2000ic,Yamagata:2005ic} atomic states, in the search
of possible anti-kaon~\cite{Baca:2000ic}, $\eta$~\cite{GarciaRecio:2002cu},
$\phi$~\cite{YamagataSekihara:2010rb}, and $D^0$~\cite{GarciaRecio:2010vt}
nuclear bound states.  Charge densities are taken from
Refs.~\cite{DeJager:1974dg,DeJager:1987qc}.  For each nucleus, we take the neutron matter
density approximately equal to the charge one, though we consider
small changes, inspired by Hartree-Fock calculations with the
density-matrix expansion~\cite{Negele:1975zz} and corroborated by pionic atom
analysis~\cite{GarciaRecio:1991wk}. All the densities used throughout
this work can be found in Table~1 of Ref.~\cite{Baca:2000ic}.  The correction
in the nuclear density to remove the finite size of the nucleon is
introduced following the scheme of
Refs.~\cite{Salcedo:1987md}\footnote{$\pi R^2$ in Eq.~(6.13) of
\cite{Salcedo:1987md} should be corrected to $\pi^2 R$.} and
\cite{GarciaRecio:1991wk}. We have also considered that in nuclei it
is necessary a finite energy, of the order of few MeV's, to extract a
nucleon. However, in nuclear matter this is not the case and
particle-hole excitations can be produced at zero energy transfer. To
improve on this deficiency, we have included in our calculation an 
average energy-gap  in the nucleon spectrum. It is used to
shift the imaginary part of the optical potential, thereby reducing the
available phase space for extracting a nucleon from the Fermi sea.
\begin{figure}[t]
\begin{center}
\includegraphics[width=1.\textwidth,angle=0]{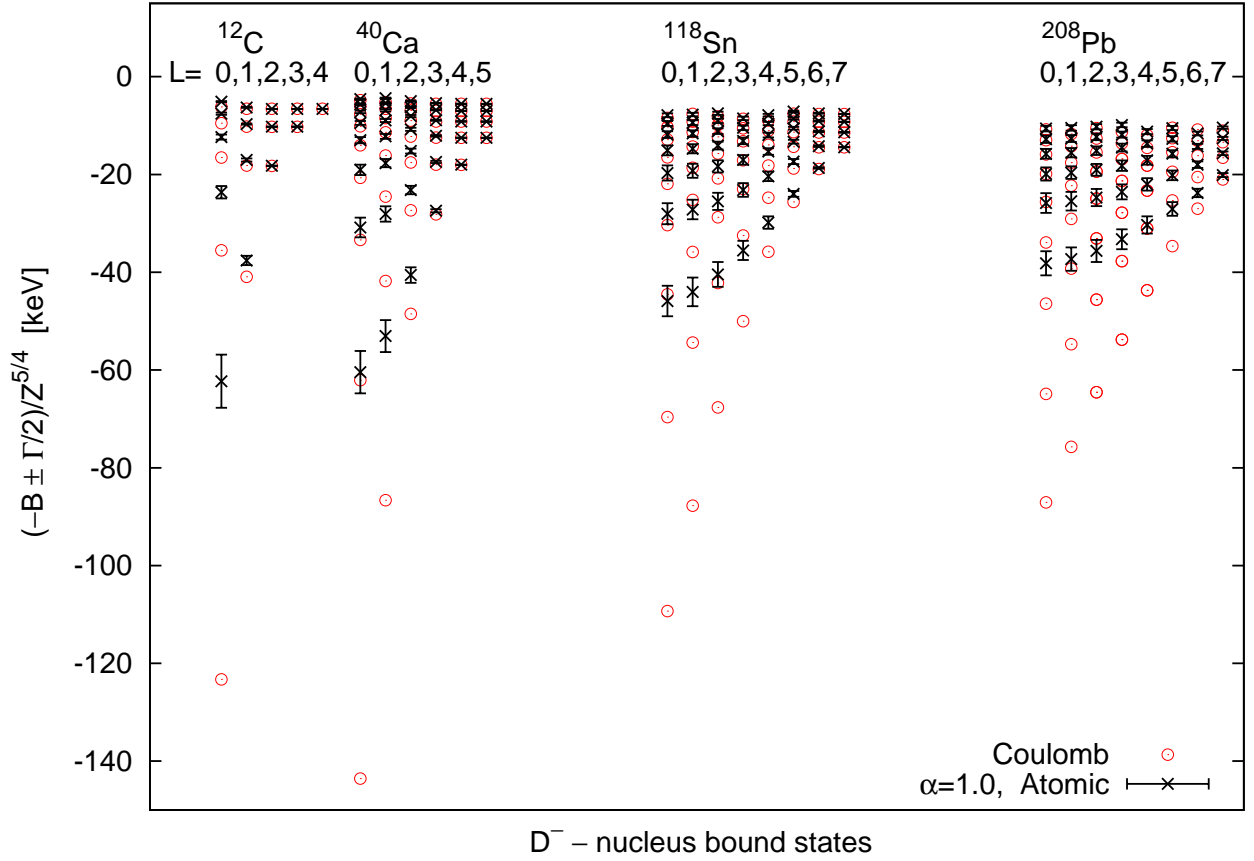}
\caption {\small $D^-$ atom levels for different nuclei and angular
  momenta. ``$\odot$'' points stand for pure Coulomb potential binding
  energies (Table \ref{tab:coul}), while ``$\times$'' symbols stand
  for the binding energies and widths of atomic levels predicted by
  the SU(8) model derived in this work (see Fig.~\ref{fig:self}), with
  $\alpha=1$ and gap $8\,\MeV$ (Table \ref{tab:a18}). The results are
  scaled down by a factor $Z^{5/4}$.}
\label{fig:levels1}
\end{center}
\end{figure}

\begin{figure}[t]
\begin{center}
\includegraphics[width=1.\textwidth,angle=0]{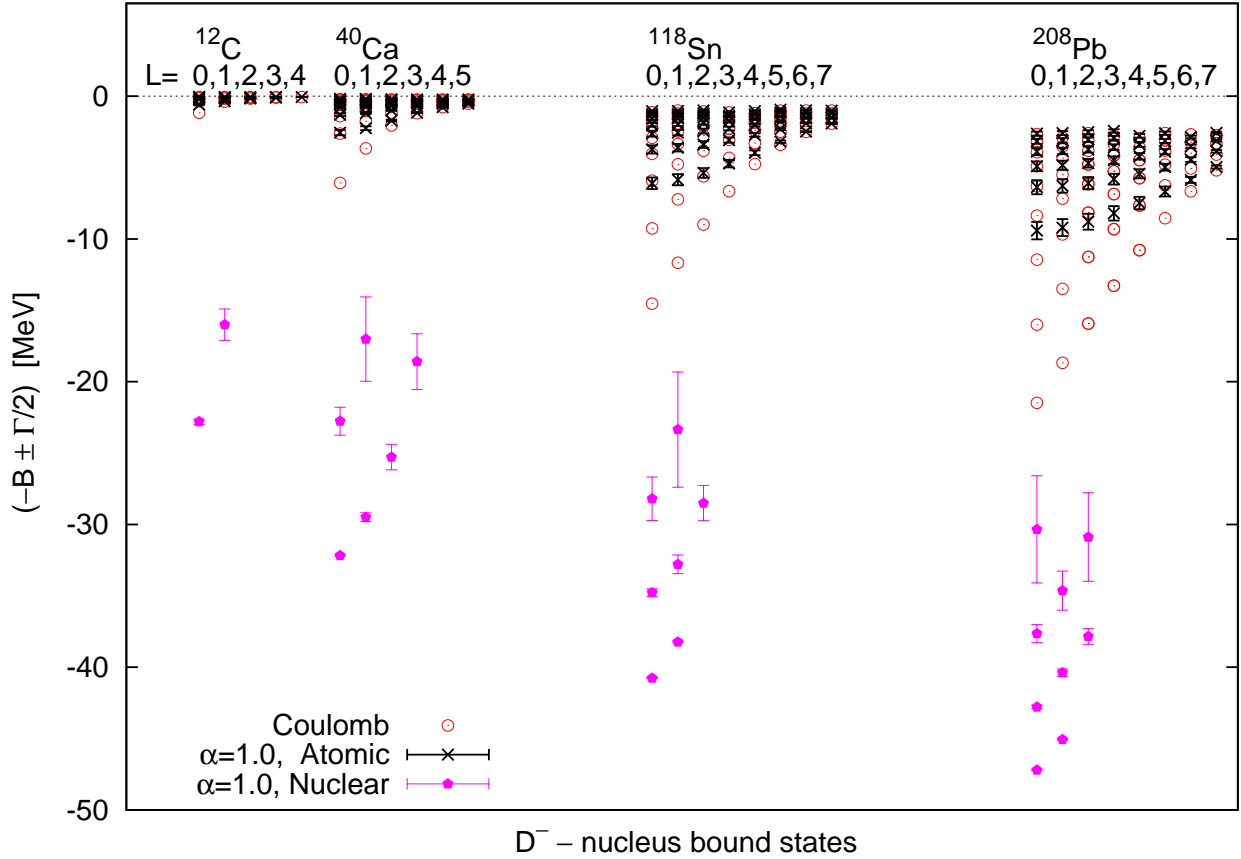}
\caption {\small Same as in Fig.~\ref{fig:levels1}, but including
  states of nuclear type (pentagons) as well. In this case no scale
  factor has been applied.}
\label{fig:levels2}
\end{center}
\end{figure}
In Tables~\ref{tab:coul}--\ref{tab:su4}, we present results for
$^{12}$C, $^{40}$Ca, $^{118}$Sn and $^{208}$Pb and for several
interactions. We have considered:
\begin{itemize}
\item[ $i)$] only Coulomb interaction, neglecting totally the nuclear
  optical potential (Table \ref{tab:coul}).
\item[ $ii)$] Coulomb interaction plus the SU(8) optical potential of
  Fig.~\ref{fig:self}, with $\alpha=1$ ($\alpha$ is defined in
  Eq.~(\ref{eq:sp})) and a gap in the nucleon spectrum of $8\,\MeV$
  (Table \ref{tab:a18}).
\item[ $iii)$] only the SU(8) optical potential with $\alpha=1$ and a
  gap of 8 MeV, thus neglecting in this case the Coulomb interaction
  (Table \ref{tab:nocoul}). This applies to $\bD^0$-nucleus states.
\item[ $iv)$] Coulomb interaction plus the SU(8) optical
potential, but with $\alpha=1.2$ or without a nucleon gap (Table
\ref{tab:Cx} with results only for $^{12}$C).
\item[ $v)$] Coulomb interaction plus the SU(4) optical potential of
  Ref.~\cite{Tolos:2007vh}, where the $\bD^* N$
  coupled-channel effects are ignored (Table
  \ref{tab:su4}).

\end{itemize}
The calculation that we deem more realistic for $D^-$ states is that
obtained by using the SU(8) model with $\alpha=1$ and with a nucleon
extraction energy (gap) of $8\,\MeV$. The predicted spectrum of
low-lying states is given in Table \ref{tab:a18} and displayed in
Figs.~\ref{fig:levels1} and ~\ref{fig:levels2}. In these figures, the
pure Coulomb levels are also shown for comparison. A salient feature
of the spectrum is the presence of two types of states: atomic and
nuclear ones.

The states of atomic type follow from distortion of the pure
Coulombian levels, they have moderate widths and they exist for all
angular momenta. For these states, the nuclear interaction is a
perturbation and their wave functions have support mainly outside of
the nucleus. As compared to the Coulombian levels, the states of
atomic type are shifted upwards, i.e., they are less bound. So
effectively, they feel a repulsive interaction.  The atomic states are
only sensible to the region of small densities and small energies, and
in this region the potential can be repulsive. (To interpret correctly
the optical potential profile in Fig.~\ref{fig:self}, it should be
taken into account that, by minimal coupling, the energy argument of
optical potential is not $q^0$ but $q^0$ increased by the local
Coulomb potential.)  Part of the repulsion comes also from the
imaginary part of the optical potential, a well known effect in exotic
atoms \cite{Krell:1971mx}. In addition, the existence of states of
nuclear type should tend to push upwards the atomic states. Yet, for
heavier nuclei, some spurious repulsion could be introduced by our
simplifying approximation of using symmetric nuclear matter in the
calculation of the optical potential\footnote{ Attending to the in
vacuum $\bD N$ $T$-matrix, since 
$(T^{(I=1,J=1/2)}-T^{(I=0,J=1/2)})(\rho_n-\rho_p)$ is negative near
threshold, the asymmetry effect is expected to be attractive for
heavier nuclei richer in neutrons than in protons.  }. As
expected, strong interaction shifts and widths become much larger for
low angular momenta and heavier nuclei. Roughly, the nuclear
interaction turns out to be significant for $L\le 1,2,5$, and $6$ for
$^{12}$C, $^{40}$Ca, $^{112}$Sn and $^{208}$Pb,
respectively. Likewise, the (strong) widths and shifts are larger for
states with lower quantum numbers due to a greater overlap of the wave
function with the nucleus. (Of course, the electromagnetic width, not
included, increases with the quantum number instead.)

On the other hand, the spectrum of the states of nuclear type lies below, and
well separated from, that of the atomic states and also from the Coulombian
levels. The gap between nuclear and atomic states ranges for 15 to $20\,\MeV$
for all nuclei, whereas the gap with the Coulomb states decreases with the
nuclear size. The nuclear states have widths ranging from few keV to several
MeV and have considerable binding energies of tenths of MeV, pointing out to a
sizable overlap of their wave function with the nucleus. The states of nuclear
type exist only for the lower angular momenta and there is only a finite
number of them that increases with the nuclear mass.  We should note that,
being the optical potential complex and energy dependent, the usual theorems
of classification of states by nodes do not apply, and so it is much harder to
guarantee that all levels in a given region of the complex energy plane have
been found.\footnote{We do not include some very wide states that overlap with
  the continuum. The fact that these states are very rare makes more
  appropriate our approximation of using only the real part of the energy as
  argument of the $\bD$ optical potential in the Schr\"odinger equation.}  An
interesting feature of the nuclear states is that their widths decrease as the
binding energies increase. (This is opposite to what happens to atomic states
regarding their strong width.) The profile of widths as a function of the
energy of the states just mimics the profile of the imaginary part of the
optical potential (see Fig.~\ref{fig:self}). The lowest states have small
widths as they fall in the tail of the imaginary part of the optical
potential. The low lying states are already inside the nucleus, so the overlap
does not increase by lowering the energy, and instead they have less
phase-space available for decay. This also explains why the widths of the
nuclear states decrease with the size of the nucleus: for larger nuclei the
ground state tends to be closer to the bottom of the potential, and hence the
available kinetic energy to knockout nucleons decreases.

In Table \ref{tab:a18} we also quote  results from Ref.~\cite{Tsushima:1998ru}
obtained in $^{208}$Pb within a quark-meson coupling model
\cite{Guichon:1987jp}. Widths were disregarded in \cite{Tsushima:1998ru}. The
numbers quoted for their model $\tilde{V}^q_\omega$ turn out to be not
very different from ours for atomic states.  Besides, one would be
tempted to say that the $1s$, $2s$ and $1p$ levels of the model
$V^q_\omega$ of Ref.~\cite{Tsushima:1998ru} match our $3s$, $4s$, and $3p$
levels of nuclear type.
\begin{figure}[t]
\begin{center}
\includegraphics[width=1.\textwidth,angle=0]{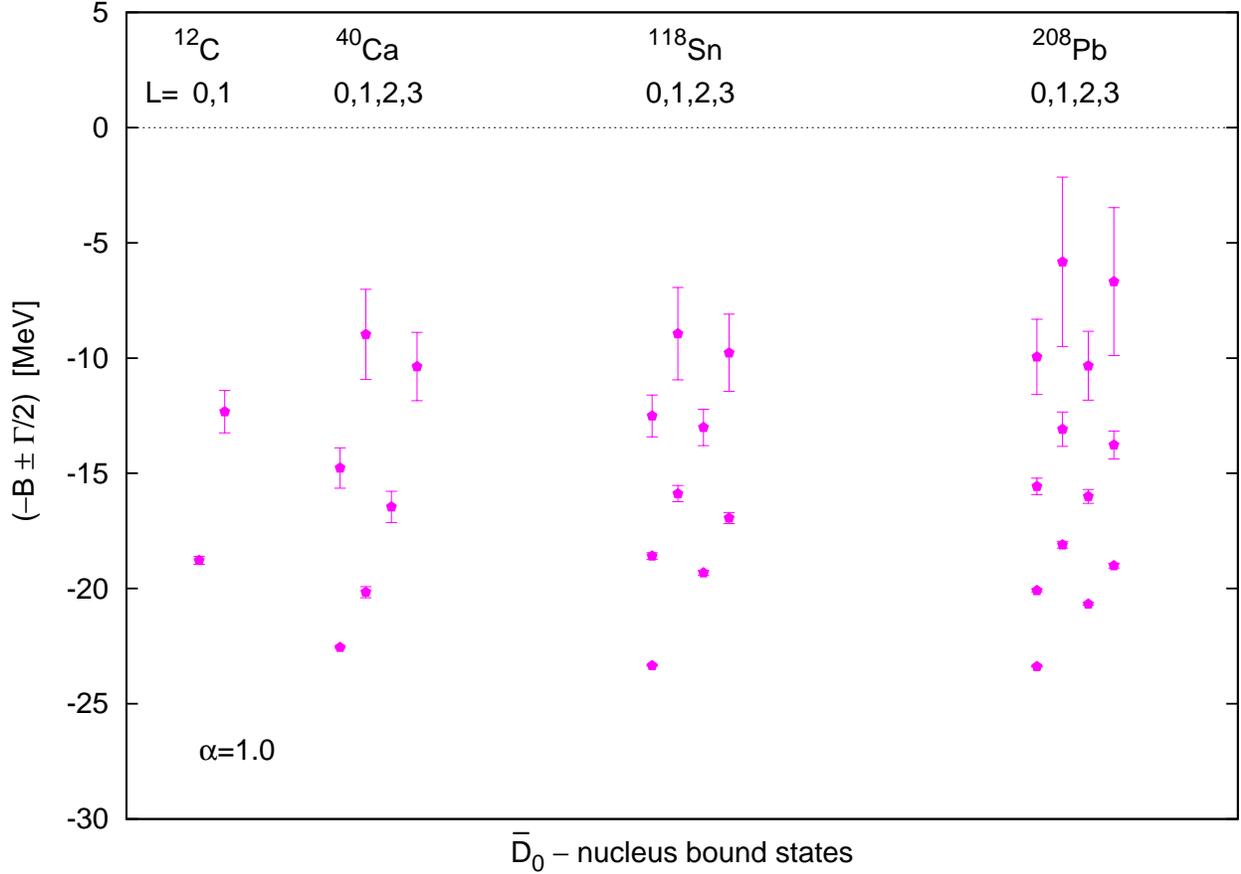}
\caption {\small $\bD^0$ nuclear levels, for different nuclei and
  angular momenta predicted by the SU(8) model derived in this work
  (see Fig.~\ref{fig:self}), with $\alpha=1$ and gap $8\,\MeV$ (Table
  \ref{tab:nocoul}).}
\label{fig:levels3}
\end{center}
\end{figure}

In Table \ref{tab:nocoul} and Fig.~\ref{fig:levels3} we show the spectrum of
$\bD^0$-nucleus bound states. This spectrum approximately matches that of the
$D^-$-states of the nuclear type. The $1p$ levels of the
two heavier nuclei are missing in the $\bD^0$ spectrum. The most likely
scenario is that those states exist but we have been unable to pin them down
due to numerical instabilities. The $\bD^0$ and $D^-$ binding energies show a
systematic difference, which can be traced back to the missing Coulomb
attraction in the $\bD^0$ case. The widths are also comparable but
systematically larger in the charged case.  This can be easily understood
since in presence of the Coulomb attraction, the binding energies are larger,
which forces the $D^-$-states to explore/be sensitive to higher nuclear
densities. In the same table we also compare with the $V^q_\omega$ model
predictions of Ref.~\cite{Tsushima:1998ru} for lead. (The model
$\tilde{V}^q_\omega$ does not produce bound states.) We find an excellent
agreement with these results for the $1s$ and $2s$ levels.

Next, we try to better understand some systematic effects that affect
to our predictions. First, we have considered the dependence of our
results on the choice of the subtraction point used to renormalize the
ultraviolet divergent loop functions.  Thus, we have re-calculated
binding energies and widths, of both atomic and nuclear levels, 
with $\alpha=1.2$. These new results are collected in
Table~\ref{tab:Cx} for carbon and can be compared with those of
Table~\ref{tab:a18} obtained with $\alpha=1.0$. From the behaviour
exhibited in Fig.~\ref{fig:self}, one might expect moderate changes,
that would lead, in general, to smaller widths and binding energies
when $\alpha$ is set to 1.2.  However, the computed $\alpha=1.2$
levels (see Table~\ref{tab:Cx}) do not always follow this pattern. The
observed deviations  are probably induced by the
strong energy dependence of the optical potential.  Though changes are
very small for atomic states with $L>0$, they become much larger for
the $L=0$ levels and also for the spectrum of nuclear states. A similar
calculation for
heavier nuclei shows that the effect is smaller, but the number of nuclear states
changes occasionally. This might be a true qualitative change in the
spectrum induced by the differences in the potentials. However, it
could be just due to the fact that some nuclear states have been
missed in the difficult search throughout the complex
plane.  For the $\bD^0$ levels, the effect of changing $\alpha$ from
1 to 1.2 is completely similar to that already noted for the states of
nuclear type in the charged case, that is, less binding and smaller
widths, and occasionally, a change in the number of levels. In the
case of the $(C=+1, S=0)$ sector, the subtraction point $\alpha$ could
be fixed in the free space by tuning the pole position of the three
star $\Lambda_c (2595)$ and $\Lambda_c (2625)$ resonances (see for
instance \cite{GarciaRecio:2008dp}). Thus in contrast with the situation here,
our previous study~\cite{GarciaRecio:2010vt} on the possible existence of 
$D^0$-nuclear bound states was free, to a large extent, of this
source of theoretical uncertainties. The lack of experimental
information on the $C=-1$ sector, however, prevents us to fix more
precisely the subtraction point used in the renormalization scheme
proposed in \cite{Hofmann:2005sw} and employed in the free space
calculation of Ref.~\cite{Gamermann:2010zz}, in which the results of this
work are based\footnote{For instance, note that for $\alpha=1$, there
exists a prominent delta-like structure in the in-medium amplitudes,
at very low densities, due to the $X(2805)$ exotic bound
state. However, it is clearly smeared out when $\alpha$ is set to 1.2,
since the $X(2805)$ baryon pentaquark is not longer bound for this
value of $\alpha$, and it becomes then a more or less narrow
resonance.}. Thus, we should take the differences between the results
displayed in Tables~\ref{tab:a18} and \ref{tab:Cx} as a hint on the nature
of the theoretical uncertainties that affect to our results. Other
sources of uncertainties will be discussed below, but among all of
them, those related to the choice of the subtraction point are
certainly the largest ones.

For instance, in the calculation we have also included in an
approximated way the effect of the existence of a non-zero gap in the
nucleon spectrum, separating nucleons in the Fermi sea from the free
ones. In Table ~\ref{tab:Cx} we  show the results without gap for
$^{12}$C.  The gap reduces the width of the states of nuclear type,
but however their binding energies are not much affected. On the other
hand, the changes in atomic states are also small. With regard to the
SU(4) model, which ignores $\bD^* N$ coupled channel effects, we
observe a repulsive interaction for the $\bD$ in nuclear matter. As a
consequence the corresponding optical potential is repulsive and
purely real in the region of interest. This model predicts $D^-$ atoms
stable under strong interactions with levels of atomic type uniformly
moved upwards in energy as compared to the pure Coulombian prediction
(see Table~\ref{tab:su4}). The repulsion is smaller than for SU(8),
presumably due to the lack of imaginary part. However, we believe the
results of Table~\ref{tab:a18} are more realistic than any of those
commented above, because neither neglecting the finite nucleon
extraction energy, nor ignoring the HQSS constraints/requirements are
approximations physically acceptable.

\section{$D^-$ atom decay modes}
\label{sec:4}

As noted in the Introduction, $D^-$ atoms\footnote{Throughout this discussion,
  ``$D^-$ atoms'' refer to all $\bD$-nucleus bound states, whether they are of
  atomic or of nuclear type.} stand out among other exotic atoms. This is also
true regarding their decay modes. Two mechanisms are available for decay,
namely, particle-hole production, $\bD \to \bD N N^{-1}$, and pentaquark
production, $\bD \to X(2805) N^{-1}$.

Let us disregard pentaquark production momentarily. A particle-hole
production mechanism is of course present in other exotic
atoms. However, in other atoms this is not the dominant decay: In
pionic atoms the non electromagnetic width comes from absorption of
the pion by two or more nucleons. In $\eta$-nucleus systems the $\eta$
carries no charge and it can be easily absorbed into particle-hole
excitations or else, it can be traded by the much lighter pion. In
$\bar{K}$-atoms the $K^-$ carries strangeness and so it cannot just be
absorbed into particle-hole excitations, but the $s$ quark can be
passed to a baryon. There is energy available for processes with
mesons in the final state, $\bar{K} \to \pi\Lambda N^{-1}$ and
$\pi\Sigma N^{-1}$, or without them, $\bar{K} \to N \Lambda
N^{-1}N^{-1}$ and $N\Sigma N^{-1}N^{-1}$ \cite{Ramos:1999ku}. Likewise, in
$D^0$-nucleus systems mesonic mechanisms, $D^0\to \pi\Lambda_c N^{-1}$
and $D^0\to \pi\Sigma_c N^{-1}$, and non mesonic ones, $D^0\to
N\Lambda_c N^{-1}N^{-1}$ and $D^0\to N\Sigma_c N^{-1}N^{-1}$, are
energetically allowed. In those decay modes, the $c$ quark is 
transferred to a baryon.

In this regard, the situation of the $D^-$ atoms is qualitatively different
from all the previous ones. Of course, the $D^-$ cannot just be absorbed into
particle-hole excitations, as was the case of pion or $\eta$.  Also, it cannot
combine with a nucleon to decay to a lighter meson-baryon channel, as happens
for $K^-$ or $D^0$, because baryons cannot carry the negative charm of the
$\bD$ and there are no lighter charmed mesons.  Put in another way, clusters
like $\bD N$ or $\bD NN$ are stable under strong decay as there are no lighter
hadronic states with same charm and baryonic quantum numbers. (Recall that the
possibility of pentaquark formation is not being considered yet.)

These remarks would also apply to an hypothetic $K$-nucleus bound
state.  However such system does not exist since the $K N$ interaction
is repulsive, as it is Coulomb for the $K^+$. On the contrary the
$D^-$ will form of necessity a bound state with the nucleus, if by no
other mechanism, through Coulomb interaction. (Even if the strong
interaction were repulsive it would vanish outside the nucleus and the
atom would be formed anyway.) So $D^-$ atoms are truly special: other
exotic atoms decay through {\em hadronic} mechanisms (to lighter
hadronic states) while $D^-$ atoms can only decay through {\em
many-body} mechanisms, e.g., $\bD N \to \bD N$, where the $\bD$ falls
to a lower level transferring energy to the nucleus.

The fact that particle-hole production is the dominant mechanism for decay has
important consequences for $D^-$ atoms, both phenomenological and theoretical.
Consider for instance the ground state of a $K^-$ atom. Although it is the
lowest atomic state, nothing prevents it from decaying to lighter hadronic
states (transferring the $s$ quark to a baryon as discussed above). On the
other hand, for the {\em ground state} of a $D^-$ atom no such lighter
hadronic state exists, so one should expect no width in this case. Put in
another way, the $\bD$ cannot go to a lower atomic state to be able to eject a
nucleon. To be precise, for the final state we should actually consider, not
the spectrum of the original atom but that of the daughter nucleus (with one
less nucleon). The difference between the two spectra is not expected to be
large, at least for not very small nuclei, and moreover, we expect the ground
state of the daughter-nucleus atom to be less bound, reinforcing the
argument. In this view, the fate of the $D^-$ atom would be to form a stable
$\bD$-nucleus bound state, which would eventually decay through weak
interaction.

From the theoretical side, the ground state argument just presented shows that
the widths, as predicted by a naive application of the $\bD$ optical
potential, tend to be overestimated for low lying states. The LDA replaces the
true discrete spectrum of the $\bD$ (in the daughter nucleus) by a continuum
of states starting from the bottom of the optical potential upwards. As the
energy of the ground state will be above that bottom, the LDA incorrectly
assigns available phase space for particle-hole decay. Of course, the same
mechanism of {\em spectral blocking} is present in the application of the LDA
to the study of other exotic atoms but this is not so crucial there because
the decay is dominated by other mechanisms which give sizable width even to
the ground state.

Another effect has to be considered as well, namely, the existence of a gap in
the spectrum of nucleons, separating nucleons in the Fermi sea from the free
ones (or from excited states, beyond the LDA). The {\em gap blocking} tends to
quench the widths from particle-hole production, as the nucleons need a
minimum energy to escape the nucleus, and so it helps to reduce or even remove
the width of low lying $D^-$ atom states. We have included such an effect in
an approximated way, just by reducing the energy argument in the imaginary
part of the optical potential by a constant amount of $8\,\MeV$.\footnote{A
  more correct procedure would be to shift only the energy of the hole line in
  Eqs.~(\ref{eq:selfd}) and (\ref{eq:selfds}), i.e., $E_N(\bm p)\to E_N(\bm
  p)-E_{\rm gap}$, but this turns out to be technically involved due to the
  presence of the $X(2805)$ state. This is a pole in the $T$-matrix which
  turns from a bound state to a resonance as the nuclear density increases.}

Up to now we have disregarded the pentaquark production mechanism.
The $X(2805)$ in the vacuum scheme of Ref.~\cite{Gamermann:2010zz} is a bound
state of $N$ and $\bD$. The binding energy is quite small, about
$1.4\,\MeV$, so this pole is very close to the $\bD N$ threshold. A
key issue is whether this state remains bound or moves to a resonance
at finite nuclear density. When Pauli blocking is enforced, the
threshold moves upwards, favoring the bound state over the resonance.
The situation changes completely when the $\bD$ optical potential is
also taken into account by means of the self-consistent procedure. In
the region of interest, the effect of the $\bD$ optical potential
turns out to be attractive. This brings the threshold downwards. In
addition, the pole in medium moves to higher energies. The net result
is that, even for a density as small as $0.1\,\rho_0$ (the lowest
density accessible by our calculation), the pole lies above the
threshold and turns into a resonance. For this density and higher, the
pentaquark would decay into $N$ and $\bD$, and this brings us back to
the particle-hole decay mechanism. Nevertheless, note that, by
continuity, there should be a critical density below which the pole is
below threshold, and so allowing the pentaquark production
mechanism. Thus this mechanism always has some contribution, however
small, to the width (even for the ground state). The situation may
also change due to gap blocking and spectral blocking, which tend to
push the threshold upwards, thereby favoring the pentaquark production
mechanism.

The presence of a pole in the $T$-matrix in the region of interest makes the
technical problem rather difficult. This has forced us to extrapolate the
optical potential from $\rho=0.1\,\rho_0$ to lower densities when needed,
without spectral blocking and with approximate gap blocking. This results in
treating the pole always as a resonance in our calculation. So, even for the
ground state, our calculation nominally attributes all widths to particle-hole
production. A more detailed treatment would provide pentaquark production as
well, below a certain critical density. In fact, for the ground state this
would be the only decay mode. Although nominally the widths we find come from
particle-hole, for low lying states there should be a genuine contribution
coming from the pole albeit distorted by the coupling to particle-hole.  This
suggests that the width computed for the ground state is a rough estimate of
the true width from pentaquark production to be obtained in a more complete
treatment without spurious particle-hole decay in the ground state.

Phenomenologically, it is important to note that the in-medium $X(2805)$ state
is produced by a bound $N$ and a bound $\bD$. Because the pentaquark formation
energy is so small, the kinetic energy released is also small and the
pentaquark remains bound in the nucleus after formation. This suggests that
after the electromagnetic cascade in the atomic levels and the subsequent
particle-hole emission cascade in the nuclear levels, the fate of the $D^-$
atom could be a pentaquark-nucleus bound state. This would be stable until
weak decay of the $\bD$ meson. Of course this is a fascinating possibility both
theoretically and experimentally.

The approximate implementation of the gap blocking and the lack of spectral
blocking suggest that the actual widths will be somewhat smaller that those
obtained here. A side effect of a smaller imaginary part would be an effective
increase in the binding of the states.

\section{Summary and conclusions}
\label{sec:5}

A self-consistent calculation of the $\bD$ self-energy has been
carried out in symmetric nuclear matter using unitarized
coupled-channel theory. The model is based on SU(8) spin-flavor
symmetry and enjoys heavy quark spin symmetry. Two renormalization
prescriptions have been used in order to estimate the systematic error
involved. We find that the presence of a bound state near the $\bD N$
threshold makes the optical potential to be strongly energy and
density dependent. In contrast with SU(4) based models, the optical
potential is mostly attractive (except for a repulsive region for low
densities near threshold, relevant for levels of atomic type) and
develops an imaginary part to particle-hole production and possibly to
pentaquark production inside the nucleus. Unlike other hadronic atoms,
no
other relevant decay mechanisms exist for the $\bD$ in nuclear matter
around threshold.

Using the local density approximation, we have computed the levels and
widths for low lying states of several nuclei, light and heavy. The
results are summarized in Figs.~\ref{fig:levels1} and \ref{fig:levels2} for
$D^-$-atoms and in Fig.~\ref{fig:levels3} for  $\bD^0$-nucleus bound states.
The spectrum presents two types of states, atomic and
nuclear, for all studied nuclei\footnote{Of course, atomic like states
do not appear in the $\bD^0$ case.}. The nuclear states
exists for lower angular momenta only. As compared to the pure Coulomb
levels, the atomic states are less bound and the nuclear ones are more
bound and may present a sizable width.

A number of approximations have been necessary in an already highly
sophisticated calculation to render it feasible. Nevertheless, this is
the first systematic study of $D^-$ atoms that accounts properly for
HQSS and for the many-body mechanisms responsible for the widths of
the states. We can draw two general conclusions from the present
work. First, that in the study of nuclear systems involving charm, it
is important to use a model fulfilling the QCD requirement of heavy
quark spin symmetry. The vector-meson partner of the $\bD$, the
$\bD^*$, has a similar mass and hence its inclusion substantially
modifies the $\bD N$ dynamics producing a non trivial structure in its
$T$-matrix near threshold. And second, the anti-quark $c$ in the $\bD$
cannot be transferred to the baryons, and in particular, a $\bD N$
pair has no open channels to decay. For this reason it has been often
assumed that the $\bD$ would not interact much with the nucleus and
could be treated as an spectator. We find that this is not the case,
and in fact a rich spectrum is obtained with sizable shifts and widths
in the states. 

The observation of the states predicted here might be feasible in the
PANDA and CBM experiments at the future FAIR facility at Darmstadt,
and it would certainly shed light to unravel the fascinating $\bD N$
dynamics, both in the free space and when the pair is embedded 
in a dense nuclear medium.

\bigskip
{\bf Acknowledgments}
We warmly thank E. Oset for useful discussions.
This research was supported by DGI and FEDER funds, under contracts
FIS2008-01143/FIS, FPA2010-16963 and the Spanish Consolider-Ingenio 2010
Programme CPAN (CSD2007-00042), by Junta de Andaluc{\'\i}a grant
FQM-225, by Generalitat Valenciana contract PROMETEO/2009/0090, and it
is part of the European Community-Research Infrastructure Integrating
Activity “Study of Strongly Interacting Matter” (acronym
HadronPhysics2, Grant Agreement n. 227431), under the Seventh EU
Framework Programme. L.T. acknowledges support from Ministerio de
Ciencia e Innovaci\'on under contract FPA2010-16963 and Ramon y Cajal
Research Programme, and from FP7-PEOPLE-2011-CIG under contract
PCIG09-GA-2011-291679, and the Helmholtz International Center for FAIR within the framework of the LOEWE  program by the State of Hesse (Germany).


\end{document}